\documentclass[a4paper,11pt]{article}
\usepackage{jcappub}
\usepackage{lineno}

\arxivnumber{2511.16967}
\title{Fast far-sidelobe modeling for centimeter to sub-millimeter astrophysical observations}

\author[a,b,1]{Oliver Jeong,\note{Corresponding author.}}
\author[a,b,1]{Jacques Delabrouille,}
\author[c,a,b]{and Michel Piat}

\affiliation[a]{CNRS-UCB International Research Laboratory, Centre Pierre Bin\'etruy, \\
IRL 2007, CPB-IN2P3, Berkeley, CA 94720, USA}
\affiliation[b]{Lawrence Berkeley National Laboratory, \\
1 Cyclotron Road, Berkeley, CA 94720, USA}
\affiliation[c]{Universit\'e Paris Cit\'e, CNRS, Astroparticule et Cosmologie, \\F-75013 Paris, France}

\emailAdd{objeong@lbl.gov}
\emailAdd{delabrouille@apc.in2p3.fr}
\emailAdd{piat@apc.in2p3.fr}

\abstract{Next-generation centimeter to sub-millimeter telescopes require exquisite control over instrumental far-sidelobe response to accurately measure faint signals like the Cosmic Microwave Background \textit{B} modes. Because existing electromagnetic modeling methods are computationally expensive, we developed a novel, diffraction-based beam modeling method for rapid and low-cost calculations. We applied this methodology to model the BICEP3 far-sidelobes and found good qualitative agreement with \textit{in situ} beam measurements. Using this validated simulated beam, we calculated the sidelobe temperature pickup for a specific observation scenario: scanning near the slopes of Cerro Toco in the Atacama Desert. This rapid, predictive framework is most valuable as a tool for optimizing instrument baffling and identifying efficient scan strategies during the conceptual design phase.}

\begin{document}
\maketitle
\flushbottom

\section{Introduction}
Since its discovery by Penzias and Wilson, observations of the cosmic microwave background (CMB) have been instrumental in establishing the concordance model of cosmology. Over the past five decades, CMB experiments have measured its anisotropies with remarkable precision, significantly advancing our understanding of the composition and evolution of the Universe. Current and forthcoming experiments are designed to search for the degree-scale parity-odd B-mode polarization pattern of the CMB, a signature imprint of primordial gravitational waves generated during cosmic inflation. The observation of such a signal would provide compelling evidence for inflation, the theorized period of exponential expansion in the early Universe, and shed light on physics at the Grand Unified Theory scales. However, this primordial B-mode signal is expected to be extremely faint and remains undetected by current CMB experiments.
\par
As next-generation CMB experiments approach unprecedented sensitivity levels, they must demonstrate exquisite knowledge and control of instrumental systematics of its small aperture telescopes (SAT) to detect primordial \textit{B} modes. 
Stray radiation pick-up through the imperfectly known sidelobes of the optical response of the instrument, also called the beam, is a significant source of instrumental systematic uncertainty in the observations.
Far sidelobe response is particularly difficult to characterize both through modeling and measurement and thus remains an active area of investigation. This challenge is especially pronounced for ground-based experiments, where contamination from ground pickup through far sidelobes is a major concern. 
\par
Modeling diffraction across large optical systems typically requires detailed electromagnetic simulations that are computationally expensive and often impractical for rapid design iteration. In this paper, we present a generalized approach to modeling the diffraction-limited response of optical systems relevant to small aperture CMB telescopes. While stray radiation can originate from various experiment-specific sources, these effects are highly dependent on the unique configuration of a given experiment and are typically mitigated through rigorous engineering. In contrast, diffraction represents a fundamental physical limit imposed by the wave nature of light interacting with the system's beam-defining apertures. By focusing on the diffraction-based framework, we provide a versatile far-sidelobe performance floor that captures the fundamental optical response of an idealized optical system. This methodology is particularly useful during conceptual design phase, as it provides a rapid means to compare different optical architectures. By isolating fundamental diffractive behaviors from situational hardware complexities, the design configurations can be rapidly explored to establish baseline performance before progressing to resource-intensive, experiment-specific modeling. 
\par
The paper is organized as follows. In Section~\ref{sec:problem}, we briefly review the problem of far sidelobes for ground-based millimeter telescopes. In Section~\ref{sec:method}, we introduce the diffraction-based method of modeling beams for these telescopes. In Section~\ref{sec:results}, we use this method to simulate the far sidelobe response of a BICEP3-like telescope and compare the results with \textit{in situ} beam measurements. Then this framework is extended to calculate the expected power pickup of the far sidelobe response of a BICEP3-like telescope observing on the slopes of Cerro Toco in the Atacama Desert of Chile. We conclude with Section~\ref{sec:conclusion}.

\section{Problem of stray radiation\label{sec:problem}}
Ground-based millimeter-wave telescopes, such as those used for detecting faint signals like the CMB \textit{B} modes or fluctuations of atomic and molecular emission lines, face a critical challenge: their detectors are highly susceptible to spurious signals from unwanted sources. The most concerning systematics include stray radiation from the ground and bright celestial objects like the Sun. To mitigate excessive contamination, observations must be constrained to ranges of allowed azimuth and elevations, ensuring that the brightest  sources of stray radiation are kept far from the telescope's line of sight. This strategy significantly degrades the observing efficiency and can still lead to reduced data quality from residual contamination, making a precise understanding of the far-sidelobe response essential for proper shielding and baffling design.

\begin{figure}[htbp]
\centering\includegraphics[width=0.75\textwidth]{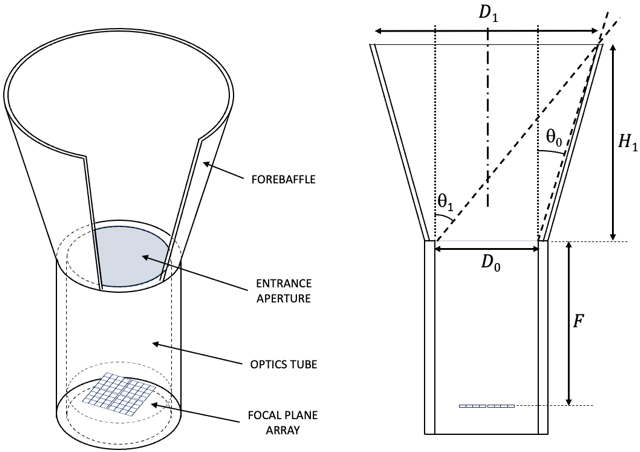}
\caption{Diagram of optical configuration used in this study with key parameters labeled, as listed in Table~\ref{tab:optical_params}.}
\label{fig:diagram}
\end{figure}

As shown in Figure~\ref{fig:diagram}, a typical CMB $B$-mode detection telescope features a main optics tube containing cold optics and a focal plane of highly sensitive detectors. To block stray radiation, these instruments incorporate several levels of shielding: a fixed ground shield is deployed first, and an additional co-moving forebaffle is placed atop the optics tube. These forebaffles are usually conical or cylindrical tubes whose inner surfaces are engineered for maximum absorption across all incident angles.

\par

The detectors are sensitive to any source of incoming radiation, meaning that their data streams register the integral of fluctuations from various forms of stray radiation. These contamination sources fall into four main categories: 1) Specular reflection of off-axis rays causing ``ghost beams'' or strong sidelobes; 2) Scattering of microwave radiation by surfaces optically coupled to the detectors, such as the vacuum window or entrance aperture; 3) Diffraction of microwave radiation by the rims of the internal shields and baffles; and 4) Thermal emission from the instrument itself, often due to absorbed sunlight being re-emitted. All these effects become problematic when strong radiation sources, like mountains or the Sun, illuminate the instrument's interior. Diffraction is particularly unavoidable and often represents the dominant sources of off-axis radiation pickup in CMB telescopes. While ray-tracing or physical optics software can be used to calculate the sidelobe response, accurate modeling requires a precise and computationally expensive definition of the entire instrument geometry, which is only feasible up to a certain level of approximation. The reader is encouraged to refer to~\cite{delabrouille} for more detailed review of sources of stray light.

\section{Diffraction-based method of modeling far sidelobe response\label{sec:method}}
\subsection{Electromagnetic principles of diffraction}
In this paper, the scalar theory of diffraction is utilized to model the propagation of light between the sky and the focal-plane of the telescope. Geometric optics is the simplest model of light propagation, in which rays travel in straight lines until refracted or reflected as described by Snell's law and Fresnel equations, respectively. When the light encounters an aperture or object, it deviates from straight-line propagation and diffracts. To account for wave-front interference that is not captured in geometric optics and accurately quantify far-sidelobes, a framework based on the Huygens-Fresnel principle is used. This principle explains this deviation by positing that every point on a wavefront acts as a source of secondary spherical wavelets, which subsequently mutually interfere and their superposition forming a new wavefront. When a wavefront encounters an aperture, only the portion of the wavefront within the transparent opening acts as a source of secondary wavelets. These wavelets then collectively form a new wavefront that bends, or diffracts, around the edges of the aperture. This is mathematically formalized by the Rayleigh-Sommerfeld diffraction integral,  

\begin{equation}
\label{eq:RS_eq}
E\left(x,y,z\right)=\frac{1}{i\lambda}\iint_{-\infty}^{+\infty}E\left(x^{\prime},y^{\prime},0\right)\frac{e^{ikr}}{r}\frac{z}{r}\left(1+\frac{i}{kr}\right)dx^{\prime}dy^{\prime}
\end{equation}
describing the propagation of the electric field in the aperture $E\left(x^{\prime}, y^{\prime}, 0\right)$ to $\left(x, y, z\right)$ where $r=\sqrt{\left(x-x^{\prime}\right)^2+\left(y-y^{\prime}\right)^2+z^2}$ and $k$ is the wavenumber $2\pi/\lambda$. The propagation of light to the far-field regime can be further approximated with the Fraunhofer diffraction equation,

\begin{equation}
\label{eq:fraunhofer_eq}
E\left(x,y,z\right)=\frac{e^{ik\left(z+\frac{x^2+y^2}{2z}\right)}}{i\lambda z}\iint_{\mathrm{Aperture}}E\left(x^{\prime},y^{\prime},0\right)e^{-i\frac{k}{z}\left(xx^{\prime}+yy^{\prime}\right)}dx^{\prime}dy^{\prime}.
\end{equation}
For an arbitrary electric field excited out of the focal plane, a combination of Eq.~\ref{eq:RS_eq} and~\ref{eq:fraunhofer_eq} can be used to model its propagation through the telescope and to the sky.

\subsection{Review of the method}
We developed an algorithm\footnote{Available at \url{https://github.com/objeong-astro/BeamModel}} to calculate the angle-dependent response, or the beam, of a telescope using \textsc{Diffractio}~\cite{diffractio}, an open-source Python package, to model the propagation of light through free-space and diffracting apertures. \textsc{Diffractio} uses a time-reversed approach, where the detector is modeled as a light source whose emission propagates through the optical system to the image plane. To emulate the focal plane excitation, the \texttt{diffractio.Scalar\_Source\_XY} module is used, enabling the flexible generation of diverse source types with customizable spatial origins and angular propagation directions. Apertures are modeled using the \texttt{diffractio.Scalar\_mask\_XY} module, which offers versatile creation of various mask shapes and types, allowing precise control over their amplitude, size, spatial placement, and tilt. For optical propagation, the \texttt{diffractio.RS} module is employed to compute the Rayleigh-Sommerfeld diffraction integral for near-field propagation, whereas the \texttt{diffractio.FFT} module handles far-field propagation via the Fraunhofer diffraction integral.
\par
In this paper, model is initiated not at the focal plane, but at the main-lobe defining entrance aperture of a telescope. At this plane, the field is Gaussian where the waist radius $w(z)$ is determined by the F-number ($N$), wavelength ($\lambda$), and the distance $z$ to the focal plane following
\begin{equation}
\label{eq:beam_waist}
w(z)=w_0\sqrt{1+\left(\frac{z}{z_R}\right)^2}
\end{equation}
where $z_R=\pi w_0^2/\lambda$ and $w_0=2\lambda N/\pi$. While this setup does not encapsulate the higher-order complexities such as antenna asymmetries or internal sidelobes, it captures the primary angular spread and illumination taper at the pupil. This provides a representative field distribution that accounts for the fundamental relationship between the focal plane and the entrance aperture. For experiment-specific applications, this tool can be modified to flexibly apply other types of illumination at different starting points. To model off-axis pixels, the beam originates from the entrance aperture at a specific angle $(\theta, \phi)$. This approach is motivated by two primary factors. First, for the refractor architectures typical of BICEP3 and SO SATs, the entrance pupil and subsequent shielding constitute the dominant diffractive apertures that dictate the far-sidelobe structure~\cite{bicep3_calib}. Second, this setup allows the model to remain experiment-agnostic for this illustrative demonstration of the method. While the internal cryogenic optics, such as lens curvature, dielectric constants, and anti-reflection coatings, shape the beam, their collective effect at the entrance pupil can be broadly summarized by a Gaussian field distribution defined by the system's F-number. This abstraction ensures the methodology remains broadly applicable to various telescope designs without requiring proprietary or experiment-specific optical prescriptions that are often not publicly available. Consequently, the internal optics are treated as a black box that transforms the feed-horn emission into a well-defined Gaussian illumination at the system's entrance pupil. However, this framework is designed to be inherently modular, allowing the user to substitute the entrance-aperture Gaussian approximation with a full propagation through the entire optics chains when such information is available. 

\begin{figure}[htbp]
\centering\includegraphics[width=0.8\textwidth]{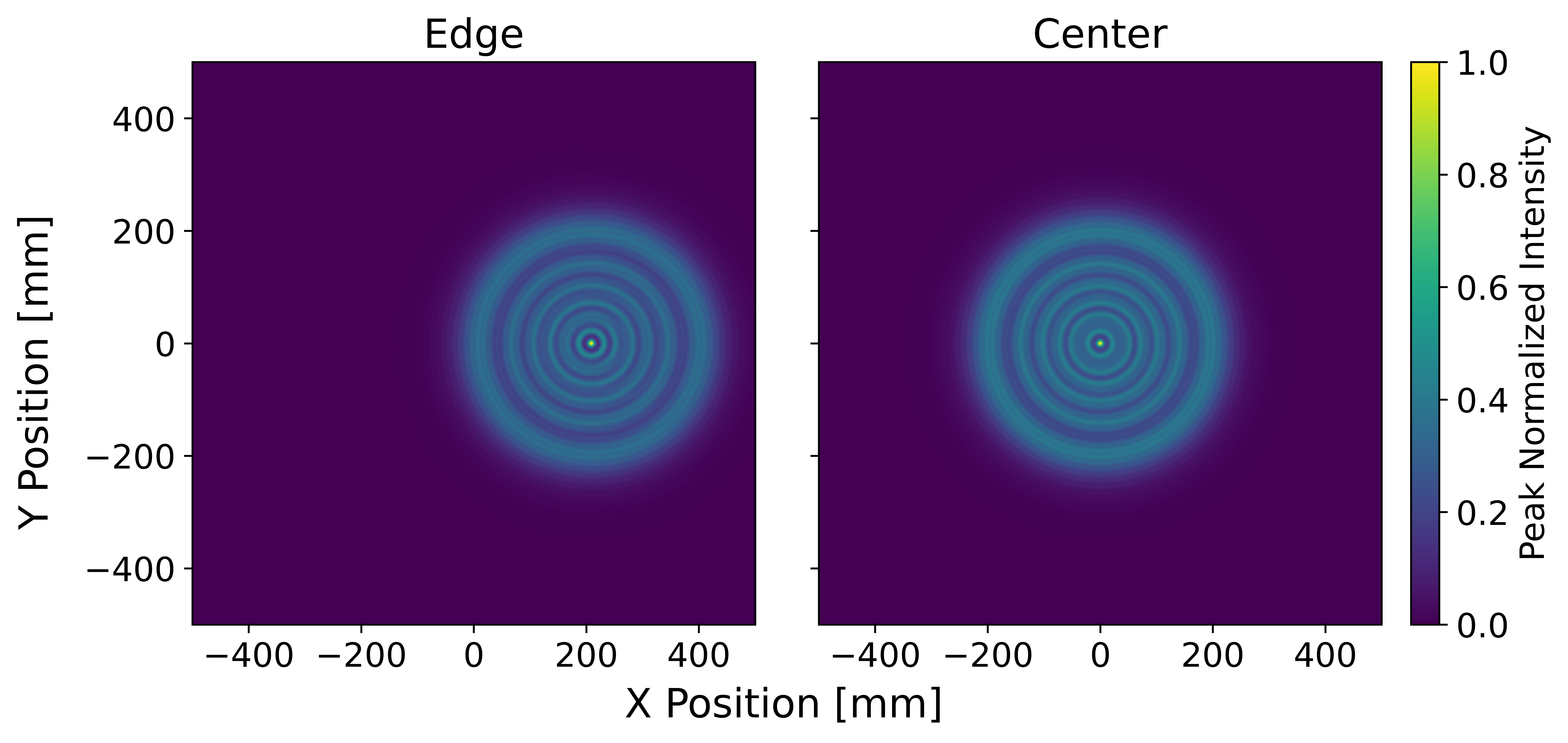}
\caption{Peak normalized beam intensity at the sky-side of the baffle for edge (Left) and center (Right) pixels at 96 GHz.}
\label{fig:beam_realspace}
\end{figure}

The Rayleigh-Sommerfeld integral propagates the tilted field to subsequent apertures, naturally accounting for the projected aperture geometry and obliquity effects experienced by the edge pixels. Figure~\ref{fig:beam_realspace} illustrates the resulting near-field intensity for both edge and center pixels. Following the final aperture, the field is propagated to the far-field. Prior to assessing the image in this region, a phase correction $C$ is applied to the field, following
\begin{equation}
\label{eq:phase_corr}
C=e^{-i\left(k_xx+k_yy\right)},
\end{equation}
where $k_x=k\sin\theta\cos\phi$, $k_y=k\sin\theta\sin\phi$ for $k$. This correction serves to align the observation coordinate system with the pixel's propagation vector, rendering the image plane orthogonal to the beam axis. The combination of the Rayleigh-Sommerfeld integral and phase correction ensures that the calculated beam profiles and suppression levels accurately represent the intrinsic diffractive response of the system as seen by the pixel, independent of zenith-aligned coordinate projections.

\section{Results\label{sec:results}}
\subsection{Simulated sidelobe maps}

\begin{figure}[htbp]
\centering\includegraphics[width=0.8\textwidth]{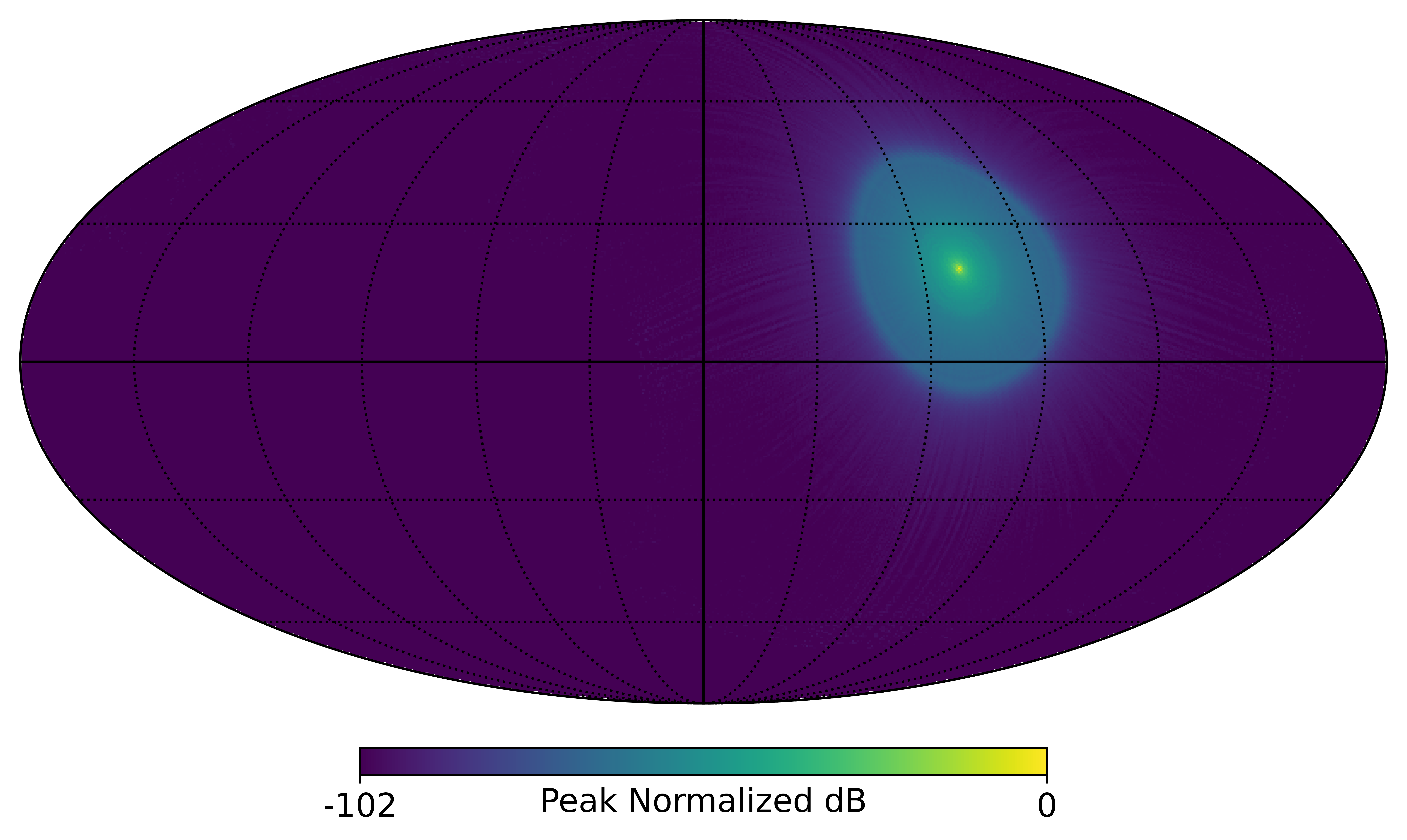}
\caption{Peak normalized beam intensity in the far-field center (Right) pixels at 96 GHz, projected onto (latitude, longitude) = (70$^{\circ}$, -110$^{\circ}$).}
\label{fig:beam_healpix}
\end{figure}

As an example application of this method, we model the beam of the BICEP3 instrument and compare against the published beam profile in Giannakopoulos \textit{et al}~\cite{bicep3_calib}. Given BICEP3's optical design, which steers the beam from a pixel on the focal plane to the center of the entrance aperture, we model its beam as originating from this aperture's center. In this model, each distinct angle of propagation from the aperture's center emulates a different pixel on the focal plane, encompassing a uniform distribution of solid angles within the telescope's defined field of view (FOV) and incorporating the appropriate beam waist. Figure~\ref{fig:beam_realspace} shows the beam intensity at the entrance of the BICEP3 forebaffle for a center and edge pixel after computing the Rayleigh-Sommerfeld diffraction integral of the fields at the entrance aperture. We multiply these near-field beams by the aperture function of the baffle and then compute their far-field diffraction integral to calculate the beam profile of a pixel of BICEP3. Figure~\ref{fig:beam_healpix} illustrates this beam profile on a HEALPix\footnote{\url{https://healpix.sourceforge.io/}}~\cite{healpix} map for a center pixel, directed towards (latitude, longitude) = (70$^{\circ}$, 110$^{\circ}$). The BICEP3 optical parameters used in this study are listed in Table~\ref{tab:optical_params}.

\begin{table}[htbp]
\caption{Optical parameters of BICEP3~\cite{B3} used to demonstrate the diffraction-based modeling capability of this method. Refer to Fig.~\ref{fig:diagram} for diagram with annotations of parameters.}
  \label{tab:optical_params}
  \centering
\begin{tabular}{ccc}
\hline
\multicolumn{2}{c}{Optical parameters} \\
\hline
$N$ & 1.6 \\
Observing band & [82.7, 109.5] GHz\\
FOV diameter & 27.4$^{\circ}$ \\
$F$ & 1078.32 mm \\
$D_0$ & 520 mm \\
$H_1$ & 1270 mm \\
$D_1$ & 1300 mm \\
\hline
\end{tabular}
\end{table}

\begin{figure}[htbp]
\centering\includegraphics[width=0.7\textwidth]{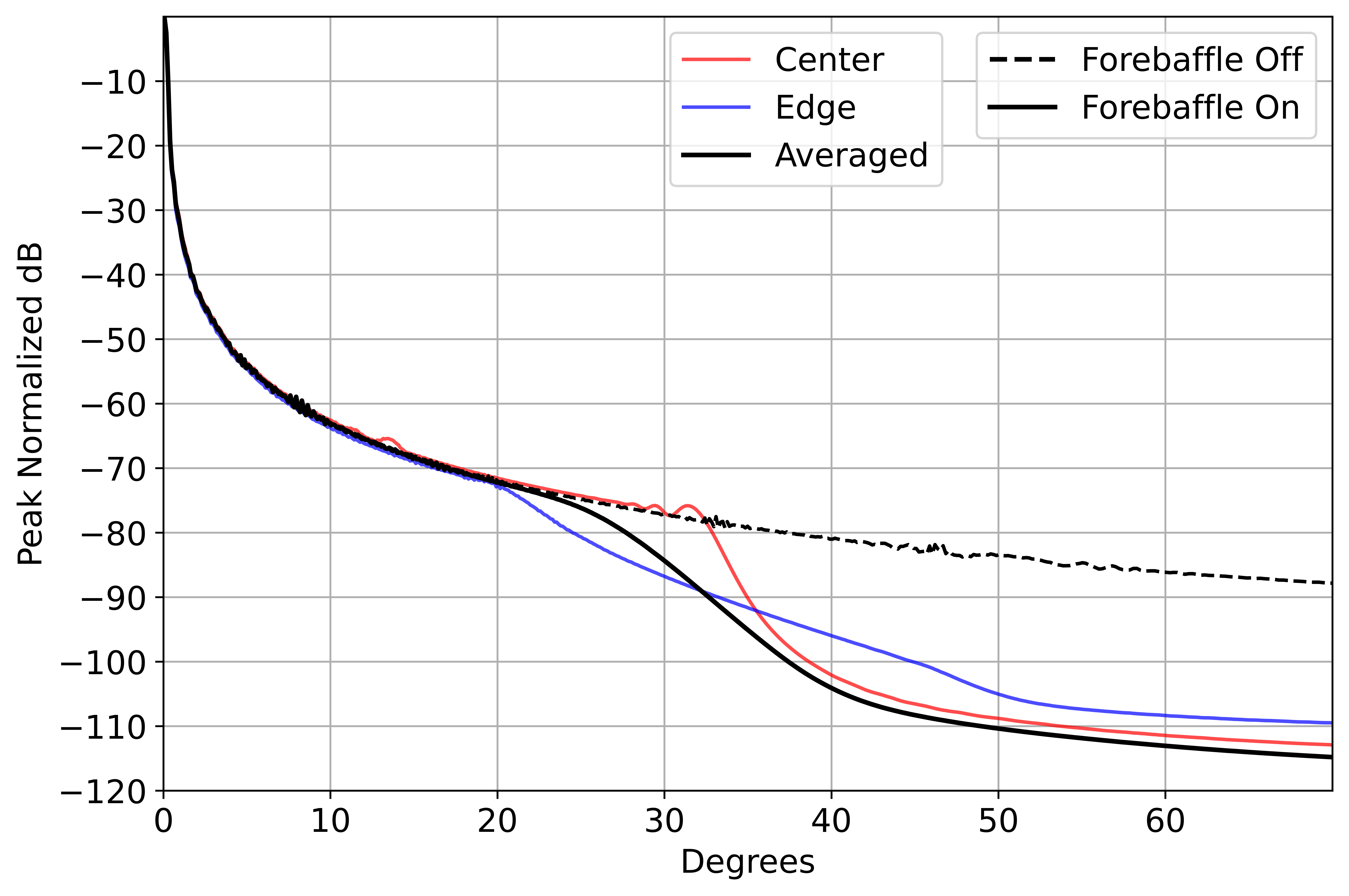}
\caption{Band- and azimuthally-averaged simulated beam profile of a BICEP3-like telescope for the center pixel (red), the edge pixel (blue), and averaged over the focal plane (black), with the forebaffle on (solid) and off (dashed). For comparison with measured beam profiles, see Figure~7 of Giannakopoulos \textit{et al.}~\cite{bicep3_calib}.}
\label{fig:beam_prof_1}
\end{figure}

The resulting beam profiles are presented in Figure~\ref{fig:beam_prof_1}, which shows the band- and azimuthally-averaged beam profile for the center pixel, the edge pixel, and the profile averaged over the focal-plane, with and without the forebaffle. As expected, the forebaffle provides strong suppression of sidelobe power of more than 20 dB for angles exceeding 36$^{\circ}$, reaching a maximum of 27 dB, for the averaged beam. Notably, the center pixel profile with the forebaffle exhibits a sharp power cutoff starting approximately at 33$^{\circ}$, which corresponds to the geometric angle subtended between the far edges of the pupil aperture and the forebaffle $(\arctan\left((D_0/2+D_1/2)/H_1\right))$. 

\par

The focal-plane-averaged profiles in Figure~\ref{fig:beam_prof_1} (black) can be compared directly with the measured BICEP3 beam profiles in Figure~7 of Giannakopoulos \textit{et al.}~\cite{bicep3_calib}. The simulation reproduces the measured main lobe and near sidelobes, with the response falling to roughly $-60$\,dB at $\sim$10$^\circ$ and $-70$\,dB at $\sim$20$^\circ$ from boresight in both cases. It also reproduces the angle, $\sim$25$^\circ$, at which the forebaffle begins to cut off the far-sidelobe response. However, a noticeable deviation is observed in the far-angle sidelobe profile. This discrepancy arises because actual sidelobe power of an operating experiment is not solely a product of diffraction but also from other sources of stray light, such as scattering on the surfaces of various optical elements. Because these contributions are highly instrument-specific, depending heavily on the exact optical elements, surface roughness, absorptive coatings, and cleanliness, modeling their contribution to the excess sidelobe power remains out of the scope of this work and users are encouraged to modify the tool for specific applications.



\subsection{Application to assess sidelobe pickup in Atacama Desert}
\begin{figure}[htbp]
\centering\includegraphics[width=0.8\textwidth]{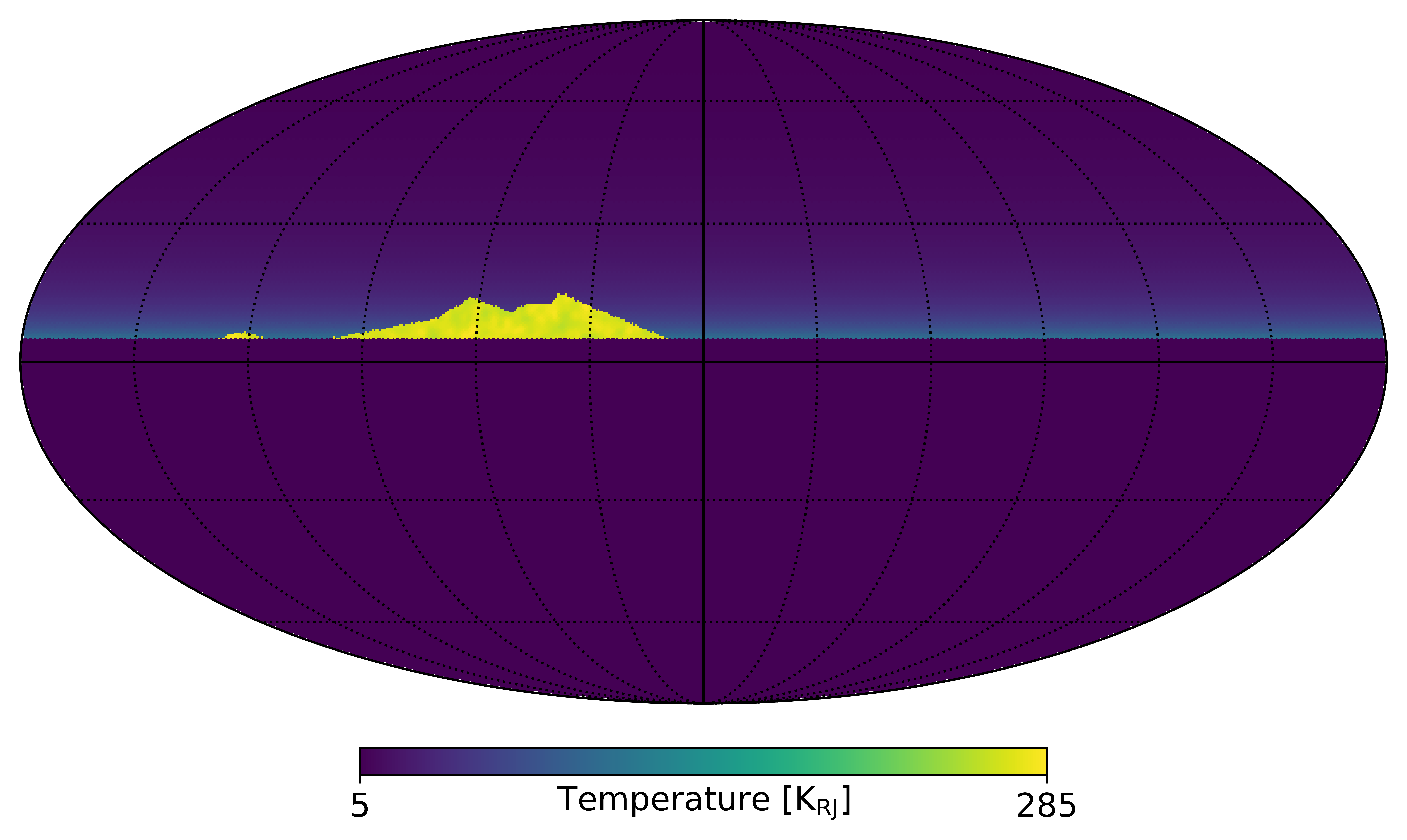}
\caption{HEALpix temperature map of the environment containing atmospheric emission from the sky at 1 mm precipitable water vapor, Cerro Toco at 270 K temperature with fluctuations of 2$^{\circ}$ scale and 5 K RMS, and ground shield perfectly reflecting sky emission at zenith.}
\label{fig:env_hpmap_lin}
\end{figure}

\begin{figure}[htbp]
\centering\includegraphics[width=0.7\textwidth]{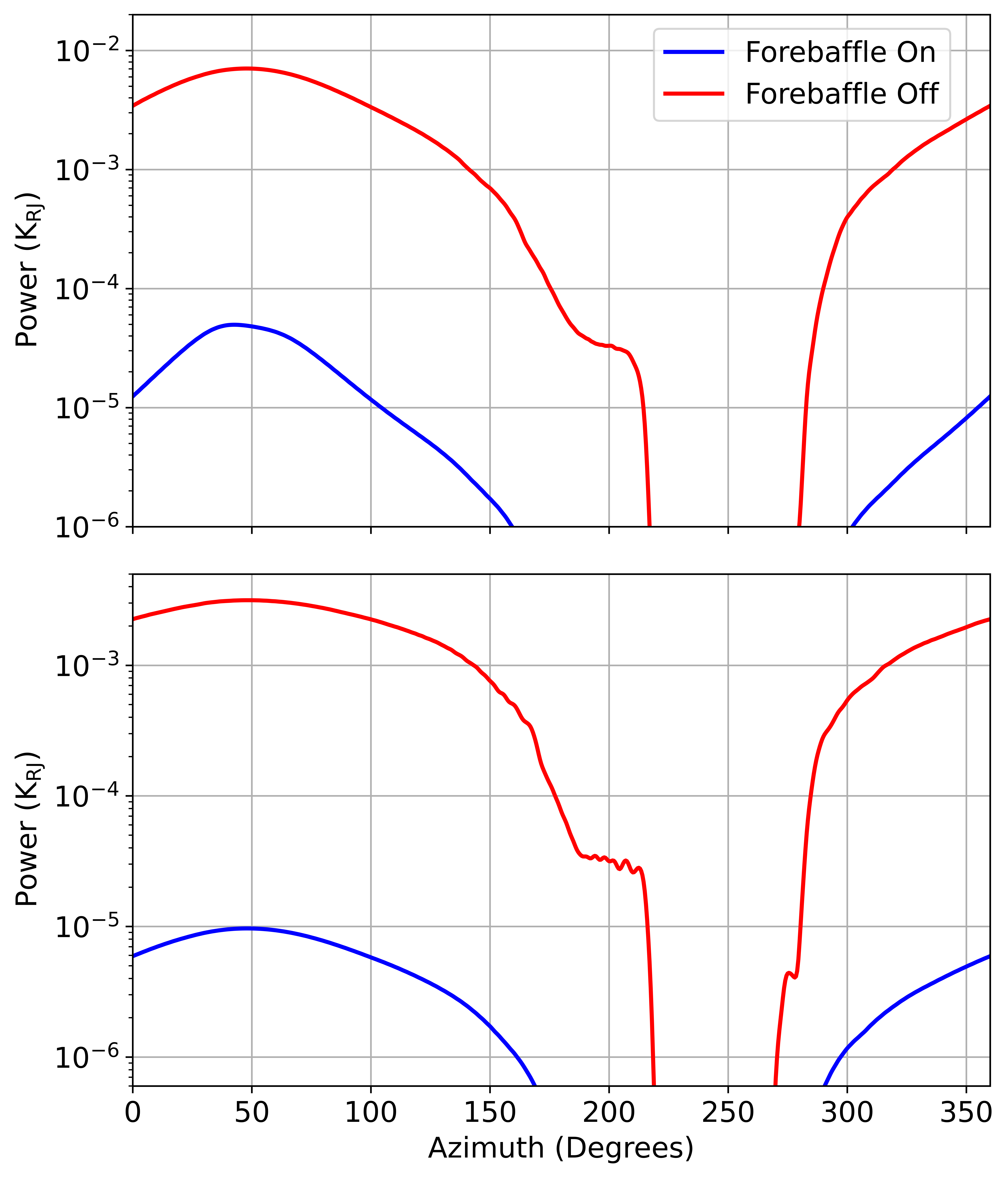}
\caption{Power pickup from the 96 GHz center pixel beam observing at 50$^{\circ}$ (top) and 65$^{\circ}$ (bottom) elevation scans.}
\label{fig:sl_power}
\end{figure}
To demonstrate the utility of this modeling framework in a semi-realistic experimental context, we simulate the sidelobe power pickup for a central pixel in a telescope with a BICEP3-like aperture observing at 90 GHz from the Atacama Desert where the small aperture telescopes of Simons Observatory are located~\cite{SO}. We emphasize that this serves as an illustrative application of the beam modeling method; while the environmental parameters are physically motivated, they are intended to show how the model can be used to estimate systematic contamination rather than provide a definitive reproduction of the Cerro Toco environment.
The total sidelobe power pickup is calculated in the Rayleigh-Jean limit by convolving the telescope's center detector beam with a HEALpix environment map, shown in Figure~\ref{fig:env_hpmap_lin}, which incorporates emission from the sky, ground screen, and the surrounding terrain. The sky temperature is modeled at 1 mm precipitable water vapor (PWV), representing the median conditions at the site, with the atmospheric path length scaling with the cosecant of the observing elevation angle, as detailed in Errard \textit{et al.}~\cite{errard}. The ground screen extends up to 5$^{\circ}$ elevation and is assumed to perfectly reflect light from the telescope vertically, setting its temperature to that of the sky at zenith. The surface of Cerro Toco is modeled at 270 K temperature with fluctuations of 2$^{\circ}$ scale and 5 K RMS. 
The total sidelobe power pickup is determined by convolving this environment map with the beam map placed at a specific scanning coordinate (see Figure~\ref{fig:beam_healpix}). From this total, we subtract the azimuthally symmetric power pickup in the absence of Cerro Toco because our primary interest lies in the spatial and temporal fluctuations caused by the terrain, rather than the static atmospheric background level. Different values of PWV would have minimal effect on this difference given the sub-dominant atmospheric background level compared to the emission from Cerro Toco. This step is functionally synonymous with the polynomial template subtraction typically applied to time-ordered data in standard CMB and other mm-wave astrophysical analysis~\cite{Ade_2022,Adkins_2026,Adachi_2022}.
\par
Figure~\ref{fig:sl_power} demonstrates the gain in sidelobe power pickup achieved using the forebaffle for scans at a constant elevation of 50$^{\circ}$ and 65$^\circ$. The forebaffle provides substantial suppression of total power pickup at all azimuthal angles. With the telescope observing directly above Cerro Toco (azimuth of $\simeq 50^{\circ}$), the telescope sees over an order of magnitude more sidelobe power with the forebaffle off.

\section{Conclusion\label{sec:conclusion}}
Next-generation CMB experiments and other centimeter to sub-millimeter experiments observing from the ground require exquisite control of systematic stray radiation pickup via instrumental far-sidelobe response. To meet this need, we have developed a diffraction-based beam modeling method that is simple and fast to directly inform the conceptual design of ground-based telescopes. When applied to model the BICEP3 far sidelobe radiation pattern, our methodology shows good qualitative agreement with the \textit{in situ} measured BICEP3 beam. This validated simulated beam was subsequently used to calculate the sidelobe pickup for a specific observation scenario near the slopes of Cerro Toco in the Atacama Desert. The simulation showed a significant difference in sidelobe pickup between the observation with the forebaffle on and the observation with the forebaffle off. This beam modeling method is most valuable as a tool for rapid comparison and optimization across various baffle geometries and instrument configurations, minimizing the need for highly detailed and computationally prohibitive full-wave simulations during the conceptual design phase. 

\acknowledgments
The authors thank colleagues at LBNL––Bobby Besuner, Shamik Ghosh, John Groh, Reijo Keskitalo, Adrian T. Lee, and Clara Vergès, for useful discussions. 
Some of the results in this paper have been derived using the healpy and HEALPix packages.

\bibliographystyle{JHEP}
\bibliography{biblio.bib}

\end{document}